\documentstyle[aps, twocolumn]{revtex}

\title{The trumping relation and the structure of the bipartite
entangled states}

\author{Sumit Daftuar\thanks{\tt daftuar@its.caltech.edu}}
\address{Institute for Quantum Information, California
Institute of Technology, Pasadena, CA 91125}

\author{Matthew Klimesh\thanks{\tt klimesh@shannon.jpl.nasa.gov}}
\address{Jet Propulsion Laboratory, California Institute of Technology,
Pasadena, CA 91109}

\newtheorem{theorem}{Theorem}
\newtheorem{lemma}[theorem]{Lemma}
\newtheorem{corollary}[theorem]{Corollary}

\begin{document}

\maketitle

\begin{abstract}
The majorization relation has been shown to 
be useful in classifying which transformations of 
jointly held quantum states are possible using local operations
and classical communication.  In some cases, a direct transformation
between two states is not possible,  
but it becomes possible in the presence of another
state (known as a {\sl catalyst}); this situation is described
mathematically by the {\sl trumping} relation, an extension
of majorization. 
The structure of the trumping relation is not nearly as well
understood as that of majorization.  We give an introduction to 
this subject and
derive some new results.
 Most notably, we show that the dimension of the required catalyst is
in general unbounded; 
 there is no integer $k$ such that it suffices to 
consider catalysts of dimension $k$ or less in 
determining which states can be catalyzed into a given state.
We also show that almost all bipartite entangled states
are potentially useful as catalysts.

\end{abstract}

\section{Introduction}
The study of quantum entanglement has received considerable 
attention in recent years, with numerous remarkable
applications including
quantum cryptography \cite{bb84,bb84p2}, quantum teleportation \cite{bbcjpw}, 
 and superdense coding \cite{bw}.  
Entanglement seems to be the essential element of such applications,
and as a result it 
has come to be viewed as a fundamental resource that
allows one to perform certain information-processing
tasks. As with any physical resource,
one wishes to measure how much entanglement is present
in a given system, and to determine under what conditions
 it is possible to 
convert one form of entanglement to another.  The problem of how
to quantify and classify entanglement is one of the basic questions
in the rapidly growing science of quantum information theory 
\cite{quicbook,preskillnotes}.

A significant advance in understanding entanglement
was made by Nielsen, who showed \cite{bignielsen} that the 
structure of the bipartite entangled states is related 
to the linear-algebraic theory of {\sl majorization} 
\cite{marshall,nielsennotes}.
We give an introduction to this subject here.
Suppose that $x = (x_1, \ldots, x_d)$ and $y = (y_1, \ldots, y_d)$
are $d$-dimensional probability vectors; in other words, their components
are nonnegative and sum to unity.  We let $x^\downarrow$ 
denote the $d$-dimensional
vector obtained by 
arranging the components of $x$ in non-increasing order: 
$x^\downarrow = (x_1^\downarrow, \ldots, x_d^\downarrow)$, where
$x_1^\downarrow \geq x_2^\downarrow \geq \cdots \geq x_d^\downarrow$.
Then we say that
$x$ is {\sl majorized} by $y$, written $x \prec y$, if the
following relations hold:

$$ \sum_{i=1}^l x_i^\downarrow \leq \sum_{i=1}^l y_i^\downarrow
    \qquad (1 \leq l < d). $$

(In fact, the theory of majorization is not limited to probability vectors.
The majorization relation can be defined as above for 
any real vectors $x$ and $y$, if we include the additional restriction
that $\sum_{i=1}^d x_i = \sum_{i = 1}^d y_i$, which
is automatically satisfied for probability vectors. For our
applications to the study of entanglement, however, $x$ and $y$ will always
be probability vectors, and we will make this assumption throughout.)

Intuitively, if $x$ and $y$ are probability vectors such that $x \prec y$, 
then $x$ describes an unambiguously more random distribution than does $y$.  
For example, in $R^2$, we have that $(0.5, 0.5) \prec (0.8, 0.2)$.  In fact,
$(0.5, 0.5)$ is majorized by every vector in $R^2$ whose components sum
to unity. 

The majorization relation defines a partial order on $d$-dimensional 
real vectors, where $x \prec y$ and $y \prec x$ if and only if
$x^\downarrow = y^\downarrow$.
 To see that majorization is not a complete relation,
consider for instance $x = (0.5, 0.25, 0.25)$ and $y = (0.4, 0.4, 0.2)$; then
$x \not \prec y$ and $y \not \prec x$.

We are now ready to state Nielsen's theorem \cite{bignielsen}:

\begin{theorem} \label{thm:nielsen}
Suppose Alice and Bob are in joint possession of a bipartite entangled quantum
state $|\psi\rangle$ which they wish to transform into another bipartite
entangled state $|\phi\rangle$ using only local operations and classical 
communication (LOCC). Let $|\psi\rangle = 
\sum_{i=1}^d \sqrt{\alpha_i}|i_A\rangle|i_B\rangle$
 be a Schmidt
decomposition of $|\psi\rangle$, and let 
$|\phi\rangle = 
\sum_{i=1}^d \sqrt{\beta_i}|i_A'\rangle|i_B'\rangle$
be a Schmidt decomposition of $|\phi\rangle$ .
Then $|\psi\rangle$ can be converted to $|\phi\rangle$ by LOCC if and only if
the vector $\alpha = (\alpha_1, \ldots, \alpha_d)$ is majorized
by $\beta = (\beta_1, \ldots, \beta_d)$.
\end{theorem}

Nielsen's theorem defines a partial order on the entangled bipartite
pure states.  If state $|\psi\rangle$ has $x$ as its vector of Schmidt
coefficients, and $|\phi\rangle$ has $y$ as its vector of Schmidt coefficients,
then we can transform $|\psi\rangle$ to $|\phi\rangle$ using LOCC
if and only if
$x \prec y$.  Because our ability to transform one state to another
depends only on their Schmidt coefficients, 
and not on the
bases, we shall abuse nomenclature and refer to any vector of
Schmidt coefficients as a ``state''.  

The above characterization of when one entangled state can be transformed to 
another is particularly helpful because the structure of the majorization
relation is relatively well understood.  For example, 
the following results are well known \cite{marshall}: 

\begin{theorem} \label{thm:majfacts}
Let $x, y \in R^d$. Then
\begin{itemize}

	\item[(a)] The following are equivalent:

	\begin{itemize}

	\item[(i)] $x \prec y$. 

	\item[(ii)]
	$\sum_{i=1}^d x_i = \sum_{i=1}^d y_i$ and for
	all $l \in \{2, \ldots, d\}$, 
	$\sum_{i=l}^d x^\downarrow_i 
	\geq \sum_{i=l}^d y^\downarrow_i$.

	\item[(iii)] $x = Dy$ for some doubly
	stochastic $d \times d$ matrix $D$.

	\item[(iv)] For every real number $t$,
	$\sum_{i=1}^d |x_i - t| \leq  \sum_{i=1}^d |y_i - t|$.

	\end{itemize}

	\item[(b)] Let $S(y)= \{x \in R^d\mid x \prec y\}$.  Then $S(y)$ is a
	convex set whose extreme points are the elements of the set
	$\{P y \mid P \mbox{ is a }d \times d \mbox{ permutation matrix}\}$.  

\end{itemize}
\end{theorem}

Jonathan and Plenio have extended Nielsen's result by
describing a phenomenon known as entanglement catalysis \cite{catpaper}.  
Suppose that $x = (0.4, 0.4, 0.1, 0.1)$ and $y = (0.5, 0.25, 0.25, 0)$.
  Then $x \not \prec y$.  Now let $z = (0.6, 0.4)$.
Then we have $x \otimes z \prec y \otimes z$.  In other words, if
Alice and Bob start only with state $x$ (by which we mean
a jointly entangled
quantum state whose Schmidt coefficients are the components of $x$),
they cannot transform it into state $y$ using LOCC\@.  But if they also
have state $z$  available, then they can turn 
$x \otimes z$ into $y \otimes z$.  So they can ``borrow'' $z$, use it 
to help turn $x$ into $y$, and ``return'' it after performing the
transformation.   We say that $z$ is 
a {\sl catalyst} for the transformation.

The phenomenon of catalysis illustrates that entanglement itself can be used
as a resource to help perform transformations of entangled states.  
One naturally wishes to know when this is possible: given $x$ and $y$, 
can we determine whether $x$ can be transformed to $y$ using LOCC in
the presence of a catalyst?  This is equivalent to asking whether there
is a probability vector $z$ such that $x \otimes z \prec y \otimes z$.  

We will adopt the terminology and notation
introduced by Nielsen \cite{nielsennotes} and say that $x$
is {\sl trumped} by $y$, written $x \prec_T y$, 
 if there exists a catalyst $z$ (of any dimension)
such that $x \otimes z \prec y \otimes z$.   For any given $y$, 
let $T(y)$ denote the set
of all $x$ such that $x$ is trumped by $y$; and for any $y$ and $z$, 
let $T(y,z)$ be the set of all $x$ such that $x \otimes z \prec y \otimes
z$.  In addition, we introduce
the following notation: for any $d$-dimensional probability vector $y$ and any positive integer $k$, 
let $T_k(y) = \{x \mid 
\exists \mbox{ a $k$-dimensional probability vector } z 
\mbox{ such that }x \otimes z 
\prec y \otimes z\}.$

  Our results will rely heavily on the fact that the trumping relation
    involves vectors with all nonnegative components.  Note that this
    is quite different from the situation with majorization, in which
    most results extend easily to vectors containing negative components.

The following facts are known about the trumping relation.
The first three are straightforward from the definitions; the others have been 
proven elsewhere \cite{catpaper,nielsennotes} . 

\begin{theorem} \label{thm:basictrump}
Let $x$ and $y$ be $d$-dimensional probability vectors, let $z$ be a
probability vector (of any dimension),  
and let $S(y)$, $T(y)$, and $T_k(y)$ be defined as above.
Then

\begin{itemize}
	\item[(a)] $x \prec y \Rightarrow x \otimes z \prec y \otimes z$.
	\item[(b)] $S(y) \subseteq T(y)$.
	\item[(c)] $T(y) = \bigcup_{k=1}^\infty T_k(y)$.
	\item[(d)] $T(y)$ is a convex set.
	\item[(e)] If $x \prec_T y$ and $y \prec_T x$, then 
		$x^\downarrow = y^\downarrow$.
	\item[(f)] If $x \prec_T y$, then $x^\downarrow_1 
		\leq y^\downarrow_1$ and $x^\downarrow_d \geq y^\downarrow_d$.
\end{itemize}
\end{theorem}

In contrast to the situation with the majorization relation, the 
mathematical structure of the trumping relation is not well understood.
One desires a
necessary and sufficient condition for determining whether $x \prec_T y$
(or alternately, to determine the elements of the set $T(y)$ for any
given $y$). Characterizing the trumping relation in this way would help
us to better understand the structure of the bipartite entangled states.
However, such a characterization is not yet known.  

In examining the trumping relation, many questions
naturally arise.  For instance, if  
$y = ({1 \over d}, \ldots,
{1 \over d})$, 
the trumping condition is (trivially) the same as the majorization
condition: $x \prec y$ if and only if $x \prec_T y$.  One wishes to know for
which $y$ this is the case.  
One also desires to know whether catalysts of arbitrarily high dimension need
be considered, in the following sense: given $y$, is it possible to find
$k$ such that $T_k(y) = T(y)$?  
These questions are among those answered in this paper.

\section{A Key Lemma}

The following lemma and its corollary will be useful to us in proving
additional results, and are also interesting in their own right:

\begin{lemma}  \label{thm:interior}
Let $x = (x_1, \ldots, x_d)$ and $y = (y_1, \ldots, y_d)$ be 
$d$-dimensional probability vectors, whose components we assume to be
arranged in non-increasing order: $x_1 \geq x_2 \geq \cdots \geq x_d$,
and similarly for $y$.  Suppose that $x \prec y$,
$y_1 > x_1$, and $y_d < x_d$.  Then $x$ is in the interior of
$T(y)$.

\end{lemma}

Note that when we say $x$ is in the interior of $T(y)$ we mean
the interior relative to the space of $d$-dimensional 
probability vectors; that
is, for any $x$ there must exist an $\epsilon$ such
that if $x'$ is a
probability vector for which $\|x'-x\| < \epsilon$ 
(in the Euclidean norm, for instance),
then $x' \in T(y)$.

We remark that the conclusion is obvious if $x$ is in the
interior of $S(y)$;
the important fact is that the result holds when $x$ is on
the boundary of $S(y)$.

{\bf \em Proof.\/}  Note that $x_d>0$.  Pick an $\alpha$ satisfying
$\alpha < 1$, $\alpha > x_1/y_1$, and $\alpha > y_d/x_d$.  Let
$k$ be an integer for which $x_1 \alpha^{k-1} < x_d$.  Now let $z$ be
the $k$-dimensional vector
\[ z = (1,\alpha,\ldots,\alpha^{k-1}). \]

(Of course $z$ is not a probability vector, but it can easily be
normalized.  For convenience in the proof, we neglect the
normalization.)

We will show that $x$ is in the interior of $T(y,z)$.  Since
$T(y,z) \subset T(y)$, this will establish the result.

Let $(y \otimes z)^{\downarrow}_i$ denote the $i$th component
of $y \otimes z$ when its components are arranged in non-increasing
order.  We will show that for $1 \leq l \leq dk-1$,
\begin{equation} \label{eq:strictmaj}
   \sum_{i=1}^{l} (x \otimes z)^{\downarrow}_i
    < \sum_{i=1}^{l} (y \otimes z)^{\downarrow}_i.
\end{equation}
    Note that since $x \otimes z$ must be majorized by $y \otimes z$,
    we already know that (\ref{eq:strictmaj}) must hold
    for $0 \leq l \leq dk$ if ``$<$'' is replaced by ``$\leq$''
    (and this fact is used later in the proof).
Showing that (\ref{eq:strictmaj}) holds for
$1 \leq l \leq dk-1$ will complete the
proof since it is then clear that any sufficiently small
perturbations to $x$ (within the probability space) will
not cause (\ref{eq:strictmaj}) to be violated for any
$1 \leq l \leq dk-1$.

For the remainder of the proof we fix $l$ as an arbitrary
integer satisfying $1 \leq l \leq dk-1$.  Consider the terms
that the left hand sum of (\ref{eq:strictmaj}) will contain.
For $1 \leq i \leq d$, let $r_i$ denote the number of these
terms which are of the form $x_i \alpha^j$, with $0 \leq j < k$.
(In case of repeated values of components of $x \otimes z$,
we regard terms with smaller $i$ to be included in the sum first.)
Note that these $r_i$ terms must be
$x_i, x_i \alpha, \ldots, x_i \alpha^{r_i-1}$, since these are
the largest of this form.  The sum (which we denote by $s_x$)
can thus be written
\begin{equation} \label{eq:xsum}
  s_x = \sum_{i=1}^d \sum_{j=0}^{r_i-1} x_i \alpha^j
\end{equation}
Note that $0 \leq r_i \leq k$
and in addition $r_1>0$ and $r_d<k$.

Consider the sum
\begin{equation} \label{eq:ysum}
  s_y = \sum_{i=1}^d \sum_{j=0}^{r_i-1} y_i \alpha^j.
\end{equation}
The terms of this sum may or may not be the $l$ largest components of
$y \otimes z$, but if $s_x < s_y$
then we are done because $s_y$ is less than or equal to the right
hand sum in (\ref{eq:strictmaj}).  The fact that $x \prec y$
implies that $s_x \leq s_y$; this follows from comparing the terms
in the sums with a fixed $j$.  Thus we need only consider the
case $s_x = s_y$.

Let $m_y$ be the minimum of the terms included in the sum
in (\ref{eq:ysum}) and let $M_y$ be the maximum of those components
of $y \otimes z$ which are {\em not\/} included in this sum.
Define $m_x$ and $M_x$ analagously.  If $M_y > m_y$ then
we are done, since the largest term not in the sum in
(\ref{eq:ysum}) can be swapped with the smallest one in the sum,
implying (\ref{eq:strictmaj}).  We assume that $M_y \leq m_y$ and
show that a contradiction will follow.

There are two cases to consider.  We first consider the
case where $r_1 < k$ (that is, $r_1 \neq k$).
Note that our current assumptions (including $M_y \leq m_y$)
imply $m_y \leq m_x$, since otherwise we would have
\[  \sum_{i=1}^{l-1} (x \otimes z)^{\downarrow}_i
    > \sum_{i=1}^{l-1} (y \otimes z)^{\downarrow}_i. \]
It follows that
\begin{equation}  \label{eq:contra1}
  m_y \leq m_x \leq x_1 \alpha^{r_1-1} < y_1 \alpha^{r_1} \leq M_y,
\end{equation}
where we have used one of our requirements on $\alpha$ as well as the
facts that $x_1 \alpha^{r_1-1}$ is in the sum in (\ref{eq:xsum})
and $y_1 \alpha^{r_1}$ is not in the sum in (\ref{eq:ysum}).
But (\ref{eq:contra1}) contradicts our assumption that
$M_y \leq m_y$, so the first case is complete.

In the other case $r_1 = k$, so $m_x \leq x_1 \alpha^{k-1}$.
But $x_1 \alpha^{k-1} < x_d$ by our choice of $k$, so we must have
$r_d > 0$.  Our assumptions imply that $M_y \geq M_x$, since
otherwise we would have
\[  \sum_{i=1}^{l+1} (x \otimes z)^{\downarrow}_i
    > \sum_{i=1}^{l+1} (y \otimes z)^{\downarrow}_i. \]
Therefore,
\[ M_y \geq M_x \geq x_d \alpha^{r_d} > y_d \alpha^{r_d-1} \geq m_y \]
by reasoning similar to that yielding (\ref{eq:contra1}).
Again our assumption that $M_y \leq m_y$ is contradicted.
Thus the proof is complete. \hfill $\Box$

\begin{corollary}  \label{cor:inttrump}
Suppose $x$ and $y$ are $d$-dimensional probability vectors,
with components arranged in non-increasing order, such that $x \prec_T y$
and $y_1 > x_1$ and $y_d < x_d$.  Then $x$ is in the interior of
$T(y)$.
\end{corollary}

{\bf \em Proof.\/}  By definition there exists a $z$ such that
$x \otimes z \prec y \otimes z$. Since $y_1 > x_1$ and
 $y_d < x_d$ we must have
    $(x \otimes z)^{\downarrow}_1 < (y \otimes z)^{\downarrow}_1$
    and $(x \otimes z)^{\downarrow}_{dk} > (y \otimes z)^{\downarrow}_{dk}$,
    where $k$ is the dimension of $z$.

We can thus apply Lemma~\ref{thm:interior} and conclude that
$x \otimes z$ is in the interior of $T(y \otimes z)$.
Since $x \mapsto x \otimes z$ is a continuous function, it
follows that $x$ is in the interior of
$\{ x \mid x \otimes z \in T(y \otimes z) \}$.
But $\{ x \mid  x \otimes z \in T(y \otimes z) \} = T(y)$, so we are done.
\hfill $\Box$

\section{When is Catalysis Useful?}

If $T(y) = S(y)$, then catalysis is of no help in producing the state $y$.
This is obviously the case when $y = (1, 0, \ldots, 0)$, for then all
vectors in $R^d$ are in both $S(y)$ and $T(y)$.  Jonathan and Plenio have
shown \cite{catpaper}
 that if $d \leq 3$ then $x \prec_T y \Rightarrow x \prec y$;
in other words,
 $S(y) = T(y)$ if $y$ is at most three-dimensional.  The following 
theorem shows that for almost all vectors $y$ of four or more dimensions,
$S(y) \not = T(y)$:

\begin{theorem} \label{thm:cathelps}

Let $y = (y_1,\ldots, y_d)$ be a $d$-dimensional probability vector
 whose components are in 
non-increasing order.  
   Then $T(y) \neq S(y)$ if and only if $y_1 \neq y_l$ and
    $y_m \neq y_d$ for some $l,m$ with $1 < l < m < d$.

\end{theorem}

This theorem says that $S(y) \not = T(y)$ if and only if $y$ has at least
two components that are distinct from both 
its smallest and largest components.

\smallskip

{\bf \em Proof.\/}
Suppose that there exist such $l$ and $m$.  Let $d_1$ be the number
of components of $y$ equal to $y_1$, and let $d_2$ be the number of
components of $y$ equal to $y_d$.  Then $d_1+d_2+2 \leq d$.
Let $x$ be the $d$-dimensional
vector whose first $d_1+1$ components are each equal to the average
of the first $d_1+1$ components of $y$, whose last $d_2+1$ components
are each equal to the average of the last $d_2+1$ components of $y$,
and which matches $y$ in any other components.  Then it is easily
checked that $x \prec y$.  In fact $x$ is on the boundary of $S(y)$
since $\sum_{i=1}^{d_1+1} x_i = \sum_{i=1}^{d_1+1} y_i$.
However, by Corollary~\ref{cor:inttrump}, $x$ is in the interior of
$T(y)$; thus $S(y) \neq T(y)$.

Conversely, assume that there are no $l,m$ such that
$l<m$, $y_1 \neq y_l$, and $y_m \neq y_d$.
Again let $d_1$ be the number of components of $y$ equal to $y_1$,
and $d_2$ the number of components equal to $y_d$.
Let $x \in T(y)$ and assume the components of $x$ are arranged
in decreasing order.  Then $x_1 \leq y_1$, so
$\sum_{i=1}^j x_i \leq \sum_{i=1}^j y_i$ for $j \in \{1,\ldots,d_1\}$.
Also $x_d \geq y_d$, so $\sum_{i=j+1}^d x_i \geq \sum_{i=j+1}^d y_i$,
and therefore
$\sum_{i=1}^j x_i \leq \sum_{i=1}^j y_i$, for
$j \in \{d-d_2,\ldots,d-1\}$.  But our assumptions imply that
$d_1+d_2 +1 \geq d$, so in fact
$\sum_{i=1}^j x_i \leq \sum_{i=1}^j y_i$ for all
$j \in \{1,\ldots,d-1\}$, and so $x \prec y$.
Thus in this case $S(y) = T(y)$.
\hfill $\Box$

In applying this theorem, it should be noted that the dimension of $y$ 
is somewhat arbitrary, as one can append zeroes to the vector $y$ 
and thereby increase its dimension without
changing the underlying quantum state.  If $y$ has at least three nonzero
components, but exactly two distinct nonzero components,
then appending zeroes will result in
a vector $y'$ such that $S(y') \not = T(y')$, although $S(y) = T(y)$.
The reason for this phenomenon is that we only consider vectors $x$
with the same dimension as that of $y$; by increasing the dimension of
$y$, we increase the allowed choices for $x$ as well.  Thus, the dimension
of the initial states $x$ under consideration may determine 
whether $S(y) = T(y)$.

\section{Catalysts of Arbritrarily High Dimension Must Be Considered}

We will now show that for most $y$, there is no $k$ such that $T_k(y) = T(y)$.
In other words, there is no limit to the dimension of the catalysts that
must be considered, in trying to determine which vectors are trumped by a given
vector $y$.  Our proof will proceed as follows: First we will show that $T_k(y)$ is a closed
set for any $k$ and all $y$, and then we will show that $T(y)$ is in general not
closed.  It follows that $T_k(y) \not = T(y)$.

The results of the previous section, and of this section, give a precise 
characterization of when $S(y) = T(y)$, and when there exists a $k$ such that
$T_k(y) = T(y)$.  While it is clear that the former situation implies the latter,
it turns out that the converse is true as well.

\begin{theorem} \label{thm:closed}

$T_k(y)$ is closed.

\end{theorem}

\emph{Proof.}
For a given $d$-dimensional probability vector $y$, let
\[ h(x,z) = \max_{1 \leq j< dk} \sum_{i=1}^j \left(
    (x \otimes z)^{\downarrow}_i -
    (y \otimes z)^{\downarrow}_i \right), \]
where $x$ and $z$ are probability vectors of $d$ and $k$
dimensions, respectively.
Observe that $h$ is a composition of continuous functions
(including the maximum of a finite set of expressions,
and the function $x \mapsto x^{\downarrow}$), and so
is continuous in $x$ and $z$.

Let
\[ f(x) = \min_z h(x,z), \]
where the minimum is over all $k$-dimensional probability vectors $z$;
this minimum exists since $h(x,z)$ is continuous in $z$ and the
minimization is over a compact set.
Observe that $x \in T_k(y)$ if and only if $f(x) \leq 0$.

Suppose now that $x \notin T_k(y)$.  Then $f(x)>\epsilon$ for
some $\epsilon>0$.  Let $x'$ be given with $\| x-x' \| < \epsilon/d$.
Let $z$ be an arbitrary $k$-dimensional probability vector,
let $j_0$ be a maximizing value of $j$
in $h(x,z)$ and $\pi$ be a permutation for which
$(x \otimes z)^{\downarrow}_i = (x \otimes z)_{\pi(i)}$ for each $i$.
Let $v$ be the $d$-dimensional vector $(\epsilon/d,\ldots,\epsilon/d)$
and note that $x'_i > x_i-v_i$ for each $i$.
We then have
\begin{eqnarray*}
   h(x',z)-h(x,z) & \geq & \sum_{i=1}^{j_0} \left(
    (x' \otimes z)^{\downarrow}_i - (x \otimes z)^{\downarrow}_i \right) \\
   & \geq & \sum_{i=1}^{j_0} \left(
    (x' \otimes z)_{\pi(i)} - (x \otimes z)_{\pi(i)} \right) \\
   & > & \sum_{i=1}^{j_0} \left(
    ((x-v) \otimes z)_{\pi(i)} - (x \otimes z)_{\pi(i)} \right) \\
   & = & -\sum_{i=1}^{j_0} (v \otimes z)_{\pi(i)} \\
   & \geq & -\sum_{i=1}^{dk} (v \otimes z)_{\pi(i)} \\
   & = & -\epsilon.
\end{eqnarray*}
Therefore $h(x',z)>0$ for all $z$, so $f(x')>0$.  We thus see
that $x' \notin T_k(y)$ for $x'$ in
a neighborhood of $x$.  Therefore $T_k^c(y)$ is open, so
$T_k(y)$ is closed.  \hfill $\Box$

\begin{theorem} \label{thm:notequal}

Let $y = (y_1, \ldots, y_d)$ be a $d$-dimensional probability vector,
with components in non-increasing order, such that $T(y) \not = S(y)$. 
Then for all $k$, $T_k(y) \not = T(y)$.

\end{theorem}

{\bf \em Proof.} 
By Theorem~\ref{thm:cathelps}, the hypothesis is equivalent to 
the existence of $l, m$
such that $1 < l < m < d$,
$y_1 > y_l$, $y_m > y_d$.
For convenience, we redefine
$l$ to be the index of the first component of $y$ that is
not equal to $y_1$, and $m$ to be the index of the last
component of $y$ that is not equal to $y_d$; clearly we
still have $l < m$.  Let
$\Delta = \min\{y_1-y_l, y_m-y_d\}$ and let $x$ be the
$d$-dimensional vector given by $x_l = y_l+\Delta$,
$x_m = y_m-\Delta$, and $x_i=y_i$ for $i \notin \{l,m\}$.
It is easily checked that $y \prec x$ but $x \not\prec y$;
therefore $x \mathrel{{\not\prec}_T} y$.
Let $w = (\frac{1}{d},\ldots,\frac{1}{d})$ and note that
$w \in S(y)$.

Suppose $T(y)$ is closed.  Since $T(y)$ is convex, the set
$\{t \in [0,1] \mid tx+(1-t)w \in T(y)\}$ is a closed
interval not containing 1, say $[0,t_0]$.  So $T(y)$
contains $t_0 x+(1-t_0)w$ as a boundary point.  But
$t_0 x+(1-t_0)w$ satisfies the hypotheses of
Corollary~\ref{cor:inttrump} and is thus an interior point of $T(y)$.
This is a contradiction, so $T(y)$ cannot be closed.  As
Theorem~\ref{thm:closed} says that each $T_k(y)$ is closed, we must
have $T_k(y) \neq T(y)$.
\hfill $\Box$

\medskip

So whenever catalysis is useful in producing $y$ (i.e., $S(y)\not = T(y)$), 
catalysts of arbitrarily high dimension must be considered.  In other words,
when $S(y) \neq T(y)$, then for any $k$ there is a $k' > k$ such that
$T_k(y)$ is a strict subset of $T_{k'}(y)$.  However, we do not know whether
increasing the catalyst dimension by one will necessarily give an 
improvement.  That is, it is unknown 
whether there is any vector $y$ and $k \geq 1$ 
such
that $S(y) \neq T(y)$ but $T_k(y) = T_{k+1}(y)$.  

\section{Which states can be catalysts?}

Another interesting question is that of which states are potentially useful
as catalysts.  If a vector $z$ is {\sl uniform}, meaning that its nonzero components
are all identical, then it is easily seen that $z$ is not capable of acting
as a catalyst: if $x \otimes z \prec y \otimes z$, then $x \prec y$ so $z$ served
no use as a catalyst. 
In \cite{nielsennotes} Nielsen conjectured 
that all nonuniform vectors are potentially
useful as catalysts. In this section, we show that this conjecture is true. 

Before we proceed, let us consider
the implications of this conjecture.  
We know already that a uniform $z$ cannot act as a 
catalyst.  A uniform $z$ with $k$ nonzero components corresponds to a maximally
entangled quantum state of Schmidt number $k$; if $k = 1$ then the state is
unentangled.
  So we have the following situation: if $z$ is a maximally 
entangled state, then $z$ cannot be used as a catalyst; but 
for any other entangled state $z$, the 
conjecture says that $z$ can serve as a catalyst.
In using entanglement as a resource, it is possible to have too much
as well as too little.  
\medskip

\begin{theorem}\label{thm:zcat}

Let 
$z = (z_1,\ldots, z_k)$ be a non-uniform probability vector.
Then there exist probability vectors $x, y \in R^4$ 
such that $x \otimes z \prec
y \otimes z$, but $x \not \prec y$. 

\end{theorem}

{\bf \em Proof.}  
We may assume without loss of generality that $z_1 \geq z_2 \geq \cdots
\geq z_k > 0$.  Define $\alpha$ and $\beta$ by the relations
$${z_1 \over z_k} = {\alpha \over \beta}$$ and
$$\alpha + \beta = 1.$$
By non-uniformity of $z$, $\alpha > \beta$.

Let $x_1 = x_2 = {1 \over 2} \alpha + {1 \over 4} \beta$, 
and $x_3 = x_4 = {1 \over 4} \beta$.  
Let $y_1 = \alpha$, let $y_2 = y_3 = {1 \over 2} \beta$, and
let $y_4 = 0$.
Let $x = (x_1, x_2, x_3, x_4)$, $y = (y_1, y_2, y_3, y_4)$.  Note
that $x \prec y$, so obviously $x \otimes z \prec y \otimes z$.
Our goal is to show that all the majorization inequalities
between $x \otimes z$ and $y \otimes z$  
are strict; in other words, for all $l \in \{1, 2,\ldots, 4k -1\}$,
\begin{equation} \label{nonuniform}
\sum_{i = 1}^l (x \otimes z)^\downarrow_i 
< \sum_{i = 1}^l (y \otimes z)^\downarrow_i.$$
\end{equation}

We will show first that the inequalities are strict when $l$ is even;
so for now, assume that $l$ is even.  There are five cases to consider.

\smallskip

\emph{Case 1:} $1 \leq l \leq k$.  We have
$$\sum_{i = 1}^l (x \otimes z)^\downarrow_i = (\alpha + {1 \over 2} \beta) 
\sum_{i = 1}^{l/2} z_i,$$ while
$$\sum_{i = 1}^l (y \otimes z)^\downarrow_i = \alpha 
\sum_{i = 1}^l z_i.$$ Thus

\begin{eqnarray*}
\sum_{i = 1}^l (y \otimes z)^\downarrow_i - 
\sum_{i = 1}^l (x \otimes z)^\downarrow_i &=& 
\alpha \sum_{i = {l/2} + 1}^l z_i - {1 \over 2} \beta
\sum_{i = 1}^{l/2} z_i \\
&=&
\sum_{i = 1}^{l/2} (\alpha z_{1/2 + i} - {1 \over 2} \beta z_i).
\end{eqnarray*}

This last quantity
is a sum of positive terms (by the definition of $\alpha$ and $\beta$),
so the inequality \ref{nonuniform} is strict.

\smallskip

\emph{Case 2:} $k+1 \leq l < 2k$.  We have  
$$\sum_{i = 1}^l (x \otimes z)^\downarrow_i = (\alpha + {1 \over 2} \beta)
\sum_{i = 1}^{l/2} z_i$$ and
$$\sum_{i = 1}^l (y \otimes z)^\downarrow_i \geq \alpha + {1 \over 2} \beta
\sum_{i = 1}^{l - k}  z_i.$$
The difference thus satisfies
$$\sum_{i = 1}^l (y \otimes z)^\downarrow_i -
\sum_{i = 1}^l (x \otimes z)^\downarrow_i \geq
\alpha\sum_{i = {l/2}+1}^k z_i - {1 \over 2} \beta
\sum_{i = l-k+1}^{l/2} z_i.$$
Note that the sums on the right hand side each contain
$k-l/2$ terms.  Since $\alpha z_i > {1 \over 2} \beta z_j$
for any $i,j$, the difference is positive, and again (\ref{nonuniform})
holds.

\smallskip

\emph{Case 3:} $l = 2k$.  In this case 
$$\sum_{i = 1}^l (x \otimes z)^\downarrow_i = \alpha + {1 \over 2} \beta$$
and 
$$\sum_{i = 1}^l (y \otimes z)^\downarrow_i \geq \alpha
+ {1 \over 2} \beta \sum_{i = 1}^{k -1 } z_i
+ {1 \over 2} \beta z_1 $$ $$= 
\alpha + {1 \over 2} \beta + {1 \over 2} \beta (z_1 - z_k) > 
\alpha + {1 \over 2} \beta,$$ so the inequality \ref{nonuniform}
is strict. 

\smallskip

\emph{Case 4:} $2k + 1 \leq l \leq 3k$.
We have 
$$\sum_{i = 1}^l (x \otimes z)^\downarrow_i = \alpha + {1 \over 2} \beta +
{1 \over 2} \beta \sum_{i = 1}^{l/2 - k} z_i$$ while
$$\sum_{i = 1}^l (y \otimes z)^\downarrow_i \geq \alpha + {1 \over 2} \beta
+ {1 \over 2} \beta \sum_{i = 1}^{l - 2k} z_i.$$  
The second quantity is clearly larger, so the inequality \ref{nonuniform}
 is strict.

\smallskip

\emph{Case 5:} $3k + 1\leq l < 4k$. 
This case is trivial because the sum for $y \otimes z$ is 1 (because there are
no more nonzero terms to be added), and the sum for
$x \otimes z$ is less than 1.

\smallskip

We have shown that (\ref{nonuniform}) holds when $l$ is even (and in the
proper range).  Now suppose $l$ is odd.  From the even cases,
it is easily verified that
\begin{equation} \label{eq:surround}
  \sum_{i=1}^{l-1} (x \otimes z)^{\downarrow}_i
  + \sum_{i=1}^{l+1} (x \otimes z)^{\downarrow}_i \\
  < \sum_{i=1}^{l-1} (y \otimes z)^{\downarrow}_i
  + \sum_{i=1}^{l+1} (y \otimes z)^{\downarrow}_i
\end{equation}
when $l \in \{1,3,\ldots,4k-1\}$.  Based on the fact that the
components of $(y \otimes z)^{\downarrow}$ are non-increasing,
$\sum_{i=1}^l (y \otimes z)^{\downarrow}_i$ is greater than
or equal to the average of the two sums in the right side of
(\ref{eq:surround}).  However,
$\sum_{i=1}^l (x \otimes z)^{\downarrow}_i$ is \emph{equal} to
the average of the sums in the left side of (\ref{eq:surround}),
since the components of $(x \otimes z)^{\downarrow}$ appear in
pairs.  We therefore see that (\ref{nonuniform}) holds when $l$
is odd.
 
Thus, the majorization inequalities are strict for all $l$ between 
1 and $4k -1$ inclusive, so
for sufficiently small $\epsilon$, $(x_1 + \epsilon, x_2 + \epsilon,
x_3 - \epsilon, x_4 - \epsilon) \otimes z \prec y \otimes z$.
However, $(x_1 + \epsilon, x_2 + \epsilon, x_3 - \epsilon, 
x_4 - \epsilon) \not \prec y$, so our theorem is proved.
\hfill $\Box$

\section{Conclusion}

While the majorization relation is a fairly well-studied subject, 
tensor-product induced majorization
(i.e., the trumping relation) is an extension of this relation about
which comparatively little is known.  
Trumping is a relatively new notion
that allows us to categorize which transformations of
entangled states are possible using only local operations
and classical communication. 
Unfortunately, given $x$ and $y$ it is not easy to
determine whether $x$ is trumped by $y$.  
And given $y$, there is no known geometric
or function-theoretic categorization of $T(y)$, 
the set of vectors trumped by $y$; this
is in contrast to the case with the majorization relation, 
where such characterizations
do exist.  In this paper we have derived a number of
results about the trumping relation, in an effort to improve our understanding
of this relation.  

Recent work has demonstrated additional applications for majorization
in quantum information theory
\cite{majsep,mixmeasure,mixedstate}.  For instance, a majorization condition
has been shown necessary for a state to be separable
\cite{majsep}; and it has
also been shown that various majorization conditions must be satisfied
by quantum systems undergoing mixing and measurement processes
\cite{mixmeasure}.  As
discoveries relating majorization to quantum information science
 are made, new
applications for the trumping relation may arise.

\section{Acknowledgements}

The authors thank Michael Nielsen for introducing us
to this subject, providing encouragement and feedback on our
results,
and generously commenting on the manuscript.
We also thank David Beckman 
for helpful discussions.

\end{document}